\documentclass[sigconf]{acmart}

\usepackage{lipsum}
\usepackage{algorithm}
\usepackage{algpseudocode}
\usepackage{fdsymbol}

\AtBeginDocument{%
  }

\copyrightyear{2025}
\acmYear{2025}
\setcopyright{cc}
\setcctype{by-nc-sa}
\acmConference[UbiComp Companion '25]{Companion of the the 2025 ACM
International Joint Conference on Pervasive and Ubiquitous Computing}{October
12--16, 2025}{Espoo, Finland}
\acmBooktitle{Companion of the the 2025 ACM International Joint Conference on
Pervasive and Ubiquitous Computing (UbiComp Companion '25), October
12 -- 16, 2025, Espoo, Finland}
\acmDOI{10.1145/3714394.3750600}
\acmISBN{979-8-4007-1477-1/2025/10}
\begin{document}

\title[Context-Adaptive Hearing Aid Fitting Advisor]{Context-Adaptive Hearing Aid Fitting Advisor through Multi-turn Multimodal LLM Conversation}

\author{Yingke Ding}
\orcid{0009-0006-3618-9492}
\affiliation{
  \institution{Tsinghua University}
  \city{Beijing}
  \country{China}
  }
\email{dyk21@mails.tsinghua.edu.cn}
\affiliation{
  \institution{University of Washington}
  \city{Seattle}
  \state{WA}
  \country{United States}
  }
\email{yingke@uw.edu}

\author{Zeyu Wang}
\orcid{0009-0007-5048-1665}
\affiliation{
  \institution{Tsinghua University}
  \city{Beijing}
  \country{China}
  }
\email{wang-zy23@mails.tsinghua.edu.cn}

\author{Xiyuxing Zhang}
\orcid{0009-0002-9337-2278}
\affiliation{%
  \institution{Tsinghua University}
  \city{Beijing}
  \country{China}
  }
\email{zxyx22@mails.tsinghua.edu.cn}

\author{Hongbin Chen}
\orcid{0009-0008-1649-8620}
\affiliation{%
  \institution{Tianjin University}
  \city{Tianjin}
  \country{China}
  }
\email{chb_22520@tju.edu.cn}

\author{Zhenan Xu}
\authornote{The corresponding authors.}
\orcid{0009-0000-0546-7580}
\affiliation{%
  \institution{Tsinghua University}
  \city{Beijing}
  \country{China}
  }
\email{xzn20@mails.tsinghua.edu.cn}

% \renewcommand{\shortauthors}{Ding et al.}

\begin{abstract}
Traditional hearing aids often rely on static fittings that fail to adapt to their dynamic acoustic environments. We propose CAFA, a Context-Adaptive Fitting Advisor that provides personalized, real-time hearing aid adjustments through a multi-agent Large Language Model (LLM) workflow. CAFA combines live ambient audio, audiograms, and user feedback in a multi-turn conversational system. Ambient sound is classified into conversation, noise, or quiet with 91.2\% accuracy using a lightweight neural network based on YAMNet embeddings. This system utilizes a modular LLM workflow, comprising context acquisition, subproblem classification, strategy provision, and ethical regulation, and is overseen by an LLM Judge. The workflow translates context and feedback into precise, safe tuning commands. Evaluation confirms that real-time sound classification enhances conversational efficiency. CAFA exemplifies how agentic, multimodal AI can enable intelligent, user-centric assistive technologies.
\end{abstract}




\maketitle

\section{Introduction}
Hearing loss is one of the most prevalent sensory impairments, constraining communication, employment, and social participation. The World Health Organization estimates that by 2050 more than 700 million people will have disabling hearing loss~\cite{who2024deafness}. For many, modern hearing aids (HAs) remain the primary intervention. Contemporary devices combine miniature microphones, amplifiers, and sophisticated digital signal-processing algorithms to improve speech intelligibility, yet they are typically fitted once in a clinic and then operate with static programs. Because everyday acoustic scenes shift rapidly, these fixed profiles often fail to deliver optimal benefit~\cite{walravens2020consistency}.

\begin{figure*}[ht!]
    \centering
    \includegraphics[width=\linewidth]{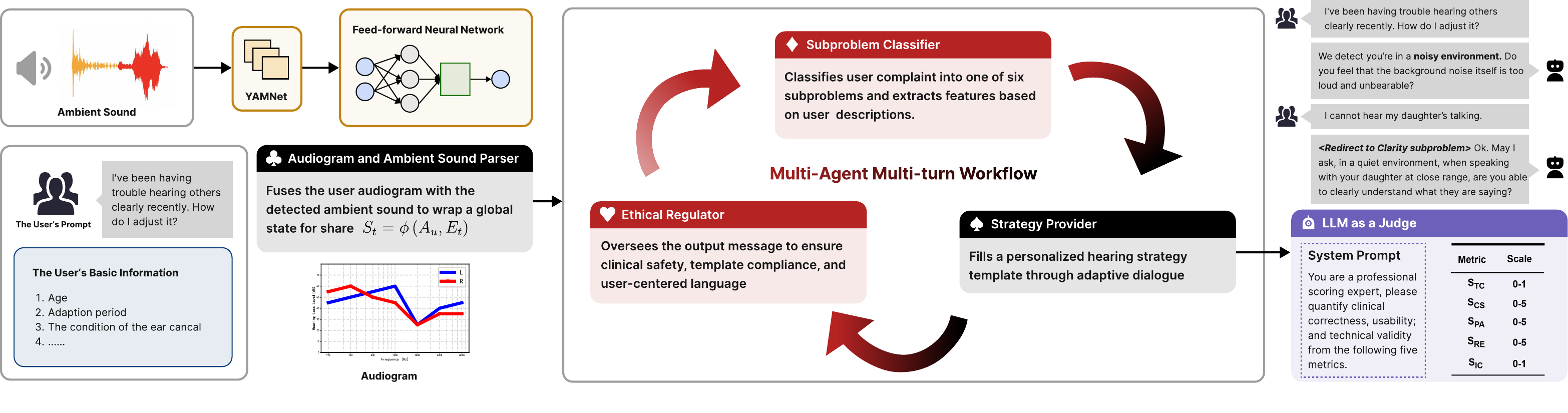}
    \caption{CAFA's system architecture. CAFA fuses lightweight audio sensing with a four-agent LLM loop for real-time, self-tuning fittings (Fig.~\ref{fig:main}). Smartphone audio is resampled to 16 kHz, embedded by YAMNet, and classified as \textit{conversation}, \textit{noise}, or \textit{quiet}, yielding the context vector $E_t$. This is concatenated with the user’s 8-band audiogram $A_u$ to form $S_t=\phi(A_u,E_t)$. Given $S_t$ and the user’s complaint, the LLM pipeline: (1) frames the context, (2) classifies the subproblem, (3) fills strategy slots, and (4) performs safety checks. The whole multi-turn conversation is evaluated by an independent LLM Judge.}
    \Description{CAFA's system architecture}
    \label{fig:main}
\end{figure*}

\textbf{Why scene awareness matters?}  Accurate, low-latency ambient-sound classification is a prerequisite for context-adaptive fitting~\cite{pasta2022measuring, wang2024dreamcatcher}. Commercial HAs now embed lightweight environment classifiers that switch programs automatically~\cite{fabry2021improving,ni2022personalization,jorgensen2023auditory}, but misclassifications, like labeling a quiet conversation as music, can degrade perception~\cite{hayes2021environmental}. Some manufacturers mitigate this by involving remote audiologists who query users and tweak settings in real time~\cite{computers7010001,korzepa2018learning}. Such human-in-the-loop services, however, are costly and do not scale. A purely automated solution must therefore (i) recognize ambient scenes reliably and (ii) remain usable even when occasional errors slip through.

\textbf{Why LLMs as complementary reasoning engines?}  Large language models (LLMs) excel at multi-step reasoning~\cite{guo2025deepseek,jaech2024openai}, nuanced question answering~\cite{lucas2024reasoning}, and agentic task execution~\cite{NEURIPS2024_90d1fc07}. Rather than replacing the scene classifier, we leverage LLMs to \emph{interpret} its output jointly with user prompts and audiometric data. This dialogue-centric approach can compensate for occasional classification errors by asking clarifying questions and proposing safe, incremental adjustments—capabilities that rule-based program switching lacks.

Commercial solutions hint at this trend. iFlytek's \emph{Spark Smart Audiologist}\footnote{\url{https://www.xunfeihealthcare.com/en/product/128.html}} offers multi-round conversational guidance driven by proprietary techniques. Signia Assistant\footnote{\url{https://www.signia.net/en-us/connectivity/signia-assistant}} uses a deep neural network to recommend adjustments that user prefers~\cite{hoydal2021assistant}. However, the systems do not exploit LLM reasoning and lacks direct access to real-time microphone data, limiting adaptability.

To bridge these gaps, we introduce CAFA, a multimodal and multi-agent LLM system that streams live audio from off-the-shelf hearing aids, ingests user audiograms and text complaints, and delivers scene-specific tuning commands through natural conversation. A built-in text-to-speech module audibly presents each recommendation, enabling users to refine their devices anywhere without sacrificing clinical rigor. We validate CAFA with objective metrics and design pilot user studies to experiment, showing the system's efficiency and perceived potentials.

Our contributions are: \textbf{1)} We propose a modular, four-agent LLM workflow that systematically translates user complaints into safe and actionable hearing aid parameter adjustments, all overseen by an LLM Judge for quality assurance. \textbf{2)} Comprehensive evaluations demonstrate that CAFA classifies ambient sound with 91.2\% accuracy and received a well-rounded conversation metrics score judged by the LLM.

\section{Methodology}
\subsection{Ambient Sound Recognition}

To enable real-time, context-aware fitting in hearing aids, we propose a low-latency, resource-efficient pipeline that classifies live audio streams on the device. First, we extract robust embeddings from a frozen, pre-trained deep audio network and then fine-tune a compact classifier tailored to different acoustic contexts.

\subsubsection{Dataset}
We construct a three-class dataset comprising \textbf{conversation}, \textbf{noise}, and \textbf{quiet} audio samples. Conversational and noise clips are sourced from the Microsoft Scalable Noisy Speech Dataset (MS-SNSD)~\cite{reddy2019scalable}, which includes 16 kHz clean speech and diverse environmental sounds. Quiet samples are self-recorded in calm indoor settings like classrooms and labs. Speech segments represent the \textit{conversation} class, while common ambient sounds (e.g., traffic, cafe noise, wind) form the \textit{noise} class, resulting in a balanced dataset for our classification task.

\subsubsection{Transfer Learning Pipeline}

We adopt a transfer learning approach for ambient sound recognition by leveraging YAMNet~\cite{google2019yamnet}, a MobileNetV1-based model pre-trained on AudioSet~\cite{45857}. Raw audio is first resampled to 16 kHz mono and processed through YAMNet to extract 1024-dimensional frame-level embeddings, which are then averaged over time to produce a clip-level feature vector. This representation is passed to a lightweight feed-forward linear classifier with one hidden ReLU layer and a softmax output, which is trained with cross-entropy loss and Adam optimizer.

\begin{algorithm}
\caption{Ambient Sound Recognition}
\label{alg:recognition}
\begin{algorithmic}[1]
\Require Audio clip $x$, pre-trained YAMNet model $M_{\text{YAMNet}}$, trained classifier $M_{\text{Classifier}}$
\Ensure Predicted class $c \in \{\text{conversation, noise, quiet}\}$

\State $x_{\text{resampled}} \gets \text{Resample}(x, 16000)$
\State $\mathbf{E} \gets M_{\text{YAMNet}}(x_{\text{resampled}})$ \Comment{Extract frame-level embeddings}
\State $\mathbf{e}_{\text{clip}} \gets \text{Mean}(\mathbf{E})$ \Comment{Aggregate embeddings}
\State $\mathbf{z} \gets M_{\text{Classifier}}(\mathbf{e}_{\text{clip}})$ \Comment{Get output logits}
\State $c \gets \text{argmax}(\text{Softmax}(\mathbf{z}))$ \Comment{Predict class}
\State \textbf{return} $c$
\end{algorithmic}
\end{algorithm}

\subsection{LLM Integration: An Agentic Workflow}
CAFA orchestrates four LLM-powered agents in a single, modular loop: \textbf{Context Acquisition} fuses the user audiogram $A_u$ with the ambient-sound $E_t$ into a global state vector $S_t=\phi(A_u,E_t)$; a \textbf{Subproblem Classifier} then maps free-text complaints to one of six canonical fitting challenges; a \textbf{Strategy Provider} conducts rapid slot-filling dialogue to turn that label into a counselling script and parameter-update fields; and an \textbf{Ethical Regulator}, all assisted by an independent LLM Judge, verifies clinical safety and policy compliance before any recommendation is shown. We engineered the system and user prompts for each agent carefully to deliver expert-level fitting guidance without domain-specific fine-tune.~\autoref{fig:main} illustrates this orchestration.

\subsubsection{Agents}\label{sec:agents}
~\\

\noindent $\clubsuit$ \textbf{Audiogram and Ambient Sound Parser.}
This agent operates at the start of a conversation by extracting the user’s hearing profile and current ambient sound context. The ambient sound is encoded as a feature vector $E_t = [e_1, e_2, \ldots, e_k] \in \mathbb{R}^k$, where each $e_j$ represents the posterior probability of a detected sound class. The user’s hearing profile is represented as an audiogram vector $A_u \in \mathbb{R}^d$, with $d = 8$ frequency bands ranging from 250 Hz to 8 kHz, and each element indicating the hearing threshold in dB HL. These two inputs are combined into a single state vector $S_t = \phi(A_u, E_t)$, which is shared across subsequent workflow stages.\\

\noindent $\vardiamondsuit$ \textbf{Subproblem Classifier.}
This agent categorizes each user complaint into one of six predefined subproblem classes: noise, distortion, clarity, loudness, blocked ears, or howl. Accurate classification is essential, as each category demands a distinct fitting strategy—for instance, directionality for noise, feedback cancellation for howl, or gain re-balancing for clarity. By transforming free-form input into structured labels, the agent (i) removes ambiguity for downstream modules, (ii) enables targeted rule retrieval, and (iii) streamlines troubleshooting. It uses an LLM guided by a structured prompt with class-specific criteria to consistently assign a single dominant label, forming a crucial step for automated, context-aware parameter adjustment.\\

\noindent $\spadesuit$ \textbf{Strategy Provider.}
This agent translates the abstract diagnostic labels assigned by the Subproblem Classifier into a concrete, step-by-step counseling and fitting plan. Its design combines a rule-based knowledge base -- hereafter the strategy book -- with an LLM-driven slot-filling dialogue manager.

\textit{Knowledge representation.}
Each subproblem is modeled as a strategy template with mandatory fields that must be populated before an actionable recommendation can be produced. We carefully picked eight fields and wrap them into a JSON object for better parsing.

\begin{table*}[ht]
\caption{Quality-control metrics used by the LLM-Judge}
\label{tab:judge_metrics}
% \small
\begin{tabular}{@{}l c c p{11.5cm}@{}} 
\toprule
\textbf{Metric} & \textbf{Symbol} & \textbf{Scale} & \textbf{Evaluation criteria} \\
\midrule
Template Compliance & $S_{\text{TC}}$ & 0–1 & Fraction of \emph{mandatory} slots that are non-null, belong to the allowed set, and satisfy all inter-slot constraints. \\
Clinical Safety & $S_{\text{CS}}$ & 0–5 & Rubric: 5=no safety issues; 3=minor risk (e.g.\ too short adaptation); 1=major risk (e.g.\ gain increase during active otitis media). \\
Personalization Adequacy & $S_{\text{PA}}$ & 0–5 & \# of distinct user-specific elements (audiogram, personal info, prior feedback) referenced. \\
Readability \& Empathy & $S_{\text{RE}}$ & 0–5 & Average of (i) readability score (Flesch $\ge$ 60 equivalence) and (ii) empathy score based on the CARE checklist. \\
Internal Consistency & $S_{\text{IC}}$ & 0–1 & Detects contradictions between narrative text and structured JSON. \\
\bottomrule
\end{tabular}
\end{table*}

\textit{Multi-turn slot filling.}
Given a classified subproblem $p$, the Strategy Provider enters a focused conversational loop with the user to collect missing slot values: 1) The agent injects the strategy template into the system prompt together with a concise explanation in the patient's language, e.g. ``When comparing male and female voices, is it harder to understand female voices, or is the clarity similar for both? Are there any significant differences for you?'', 2) At each turn $\tau$, the LLM selects the most informative unanswered slot $s_\tau$ by maximizing an information-gain heuristic

\begin{equation}
    s_\tau=\arg \max _{s \in S_{\text {empty }}} \frac{H(\operatorname{allowed}(s))}{\left|S_{\text {empty }}\right|}
\end{equation}
where $H$ is the categorical entropy over the slot's allowed values. The question is phrased with contextual examples taken from the knowledge base to avoid user confusion.

\paragraph{Conflict Resolution}
If newly supplied information violates any domain rule, the agent triggers an \emph{exception-repair sub-dialogue}, asking targeted follow-up questions that prompts the user to re-answer the specific question again.

The loop terminates when all mandatory slots are filled or a pre-set turn limit (in our case, 10) is reached.\\

\noindent $\varheartsuit$ \textbf{Ethical Regulator}
Before committing, we apply this agent to review the script and JSON output for clinical safety, template compliance, and user-centered language, ensuring a final, validated strategy is safely provided.

\subsubsection{LLM as a Judge}
After the LLM system emits a counselling script and its accompanying JSON payload, a dedicated \textbf{LLM Judge} conducts an automatic quality-control pass. Similar to~\cite{xie2024psydt}, we propose five orthogonal metrics to quantify clinical correctness, usability, and technical validity as shown in~\autoref{tab:judge_metrics}.

\section{Experiments}

\subsection{Experimental Setup}
\subsubsection{Hardware and Software}
Experiments were conducted using Bluetooth-LE hearing aids paired with an iPhone 14 Pro. Ambient audio was recorded at 48 kHz/24-bit and downsampled to 16 kHz/16-bit for on-device processing. CAFA agents were deployed using Dify~\cite{difygithub2025} v1.5.0, with GPT-4.1 powering the Parser and Strategy Provider, and GPT-4o-latest used for the Ethical Regulator and Classifier (temperature 0.7, max 8k tokens). The LLM Judge used OpenAI’s o3 model with the same settings to avoid data leakage. All prompts were written in Chinese to match audiologist practice and ensure linguistic consistency. %~\autoref{fig:clarity_strategy} is an example audiologist's strategy for clarity subproblem and~\autoref{fig:dify_strategy} is the Dify software screenshot for CAFA.

\subsubsection{Synthetic Simulated Fitting Workflow}
To test the pipeline before human trials, we scripted 130 synthetic sessions. A scenario generator sampled audiometric profiles from the ISO 7029 distribution (mild, moderate, severe) and combined them with six canonical problem types as described in Section~\ref{sec:agents}. We prompt-engineered an additional GPT-4o based LLM to act as a virtual user, conditioned on the sampled background information. Ground-truth fitting actions for each scenario were determined by two licensed audiologists to provide gold-standard labels for the evaluation.

\subsubsection{User Study}
A within-subjects user study is under Institutional Review Board (IRB) review. We have pre-registered to recruit 10 adults with mild-to-moderate hearing loss (pure-tone average $\le 55 \text{dB HL}$). Participants will complete two counter-balanced conditions: CAFA (four LLM agents) and Baseline (traditional audiologist-user conversation). Primary outcomes are task completion time, NASA-TLX cognitive load, and System Usability Scale (SUS). Secondary outcomes include self-reported listening benefit and conversational turns.

\subsection{Main Results}
\subsubsection{Ambient-Sound Classification Accuracy}
Our lightweight classifier achieves an \textbf{overall accuracy of 91.2\%}. Macro-averaged precision, recall, and F1 score are 0.914, 0.903, and 0.908, respectively. Most residual confusion occurs between \textit{conversation} and \textit{noise} scenes containing background chatter.

\subsubsection{LLM Judge}
We asked the judge to verify whether the strategy suggested by the action-planning agent adhered to the metrics provided in~\autoref{tab:judge_metrics}. The agent achieved average scores of 0.73 for $S_{\text{TC}}$, 4.25 for $S_{\text{CS}}$, 3.07 for $S_{\text{PA}}$, 4.76 for $S_{\text{RE}}$, and 0.95 for $S_{\text{IC}}$.

\subsection{Ablation Study}
For the synthetic workflow experiment, removing the ambient-sound parser agent increased dialogue length from an average of $6.7$ to $9.4$ turns. Qualitative analysis revealed that missing sound scenes led the reasoning agent to repeatedly query for context, underscoring the parser's contribution to conversational efficiency. For the other agents, removal would render the system incomplete; therefore, ablation was not conducted on them.

\subsection{Discussion}
Results indicate that CAFA accelerates fitting, and maintains guideline adherence by explicitly injecting scene awareness before language reasoning. Nonetheless, several limitations remain: the sound dataset is modest and English-centric, potentially biasing performance in non-Western acoustic scenes, and synthetic users cannot fully represent human behavior. Future work will enlarge the corpus with multilingual data, and investigate adaptive prompts that personalize both acoustic and linguistic priors to individual users.

\section{Conclusion}
In this paper we presented CAFA, a context-adaptive hearing aid fitting advisor that awares ambient sound within multi-agent LLM workflows to deliver personalized, conversational hearing aid fitting for hearing aid users. Our experiments demonstrate that CAFA achieves high ambient sound classification accuracy, enables efficient and effective fitting dialogues, and aligns closely with audiologist best practices through its LLM-driven reasoning.

\begin{acks}
This work is supported by the National Key Research and Development Program of China under Grant No.2024YFB4505500 \& 2024YFB4505503, the Natural Science Foundation of China under Grant No. 62472244, Qinghai University Research Ability Enhancement Project (2025KTSA05), the Beijing Key Lab of Networked Multimedia, the Tsinghua University Initiative Scientific Research Program, and the Undergraduate Education Innovation Grants, Tsinghua University.

The mentors of this work is Yuntao Wang and Yuanchun Shi.

\end{acks}

\bibliographystyle{ACM-Reference-Format}
\bibliography{cafa}

\end{document}